\documentclass[sigconf,nonacm]{acmart}
\setcopyright{none}
\renewcommand\footnotetextcopyrightpermission[1]{}
\usepackage{fontawesome5}
\usepackage{booktabs}
\usepackage{pifont}
\usepackage{array}
\usepackage{multirow}
\usepackage{makecell}
\usepackage{float}
\usepackage{caption}
\usepackage{adjustbox}
\usepackage[table,xcdraw,dvipsnames]{xcolor}
\usepackage{enumitem}
\usepackage{hyperref}

\usepackage[most]{tcolorbox}
\newtcolorbox{claimbox}[1][]{
  enhanced,
  breakable,
  colback=gray!5,
  colframe=gray!40,
  boxrule=0.5pt,
  arc=2mm,
  left=2mm, right=2mm,
  top=3mm, bottom=1mm,
  fonttitle=\bfseries,
  coltitle=black,
  colbacktitle=gray!20,
  attach boxed title to top left={xshift=2mm, yshift=-1mm},
  boxed title style={
    sharp corners,
    colframe=gray!40,
    colback=gray!20,
    size=small,
  },
  title={\ifstrempty{#1}{}{#1}},
}

\definecolor{chatgptgreen}{RGB}{16,163,127}
\definecolor{geminiblue}{RGB}{66,133,244}
\definecolor{claudeorange}{RGB}{204,93,44}
\definecolor{teal}{RGB}{16,163,127}
\definecolor{softgreen}{RGB}{60, 130, 70}
\definecolor{softred}{RGB}{190, 70, 60}
\definecolor{softyellow}{RGB}{240, 198, 116}
\definecolor{softblue}{RGB}{138, 173, 208}
\definecolor{softgray}{RGB}{200, 200, 200}
\definecolor{colorrorange}{HTML}{FF914E}
\definecolor{colorrred}{HTML}{E867A7}
\definecolor{colorrblue}{HTML}{2525BA}
\definecolor{colorrpurple}{HTML}{947FF2}
\definecolor{colorrgray}{HTML}{726869}
\definecolor{colortextblue}{HTML}{004C99}
\definecolor{colortextgreen}{HTML}{1F6D42}
\definecolor{colortextorange}{HTML}{E46725}

\newcommand{\softgreen}[1]{\textcolor{softgreen}{#1}}
\newcommand{\softred}[1]{\textcolor{softred}{#1}}

\newcommand{\rorange}[1]{\textcolor{colorrorange}{#1}}

\newcommand{\rgray}[1]{\textcolor{colorrgray}{#1}}

\definecolor{SearchBand}{RGB}{228,244,234}     % light green tint
\definecolor{UnderstandBand}{RGB}{232,238,250} % light blue tint
\definecolor{HeadBand}{RGB}{36,36,36}          % dark header text color (optional)

\newcommand{\redone}[1]{$_{\color{BlueGreen}\uparrow #1}$}
\newcommand{\red}[1]{$_{\color{RedOrange}\uparrow #1\%}$}

\AtBeginDocument{%
  }

\acmConference[]{}{}{}
\acmBooktitle{}
\acmYear{}
\acmDOI{}
\acmISBN{}

\begin{document}

% \title{AutoSearchBench: Automatically Generated Benchmark for LLM-based Scholarly Search Evaluation}
% \title{TrustResearch: Benchmarking the Reliability of LLMs in Paper Search and Reading}
\title{PaperAsk: A Benchmark for Reliability Evaluation of LLMs in Paper Search and Reading}

\author{Yutao Wu}
\affiliation{%
  \institution{Deakin University}
  \country{}
}
\email{oscar.w@deakin.edu.au}

\author{Xiao Liu}
\affiliation{%
  \institution{Deakin University}
  \country{}
}
\email{xiao.liu@deakin.edu.au}

\author{Yunhao Feng}
\affiliation{%
  \institution{Fudan University}
  \country{}
}
\email{yunhaofeng921@gmail.com}

\author{Jiale Ding}
\affiliation{%
  \institution{Fudan University}
  \country{}
}
\email{22307140024@m.fudan.edu.cn}

\author{Xingjun Ma}
\authornote{Corresponding author.}
\affiliation{%
  \institution{Fudan University}
  \country{}
}
\email{xingjunma@fudan.edu.cn}

\begin{abstract}
Large Language Models (LLMs) increasingly serve as research assistants, yet their reliability in scholarly tasks remains under-evaluated. In this work, we introduce \textbf{PaperAsk}, a benchmark that systematically evaluates LLMs across four key research tasks: citation retrieval, content extraction, paper discovery, and claim verification. We evaluate GPT-4o, GPT-5, and Gemini-2.5-Flash under realistic usage conditions—via web interfaces where search operations are opaque to the user.  Through controlled experiments, we find consistent reliability failures: citation retrieval fails in 48–98\% of multi-reference queries, section-specific content extraction fails in 72–91\% of cases, and topical paper discovery yields F1 scores below 0.32, missing over 60\% of relevant literature. Further human analysis attributes these failures to the uncontrolled expansion of retrieved context and the tendency of LLMs to prioritize semantically relevant text over task instructions. 
Across basic tasks, the LLMs display distinct failure behaviors: ChatGPT often withholds responses rather than risk errors, whereas Gemini produces fluent but fabricated answers.
To address these issues, we develop lightweight reliability classifiers trained on PaperAsk data to identify unreliable outputs. 
PaperAsk provides a reproducible and diagnostic framework for advancing the reliability evaluation of LLM-based scholarly assistance systems.
\end{abstract}





\maketitle

\section{Introduction}

Researchers increasingly rely on LLM-powered tools such as Deep Research, Gemini, and specialized systems like Consensus~\cite{openai2025deepresearch,google2025gemini,consensus2025} to find papers, extract information, and verify relevance. When these systems cite sources, they appear capable of integrating search to generate accurate responses. 
However, prior studies show that LLM-generated references are often incorrect~\cite{algaba2025deep,Marcus2025AIRefs}, revealing persistent issues of fabrication and unreliable sourcing. It therefore remains unclear whether modern, search-augmented LLMs truly overcome these limitations—a question this work seeks to answer.

The shift from parametric-only LLMs to search-augmented systems has introduced new capabilities. 
Modern LLMs now support extended context windows, integration with the Model Context Protocol (MCP), and enhanced safety alignment to reduce hallucinations~\cite{shi2025tool, yu2025memagent, ma2025safety}. 
These advances suggest that modern LLMs may alleviate earlier limitations, yet systematic evaluation and empirical evidence to validate this assumption remain underexplored.

\begin{figure}[!t]
    \centering
    \includegraphics[width=\linewidth]{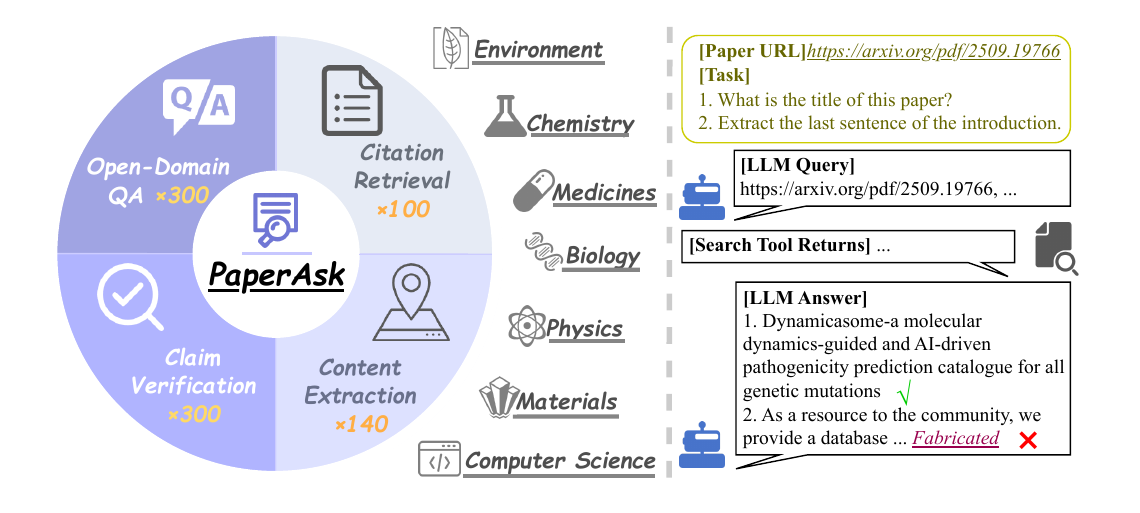}
    \caption{Overview of \textbf{PaperAsk} (4 tasks covering 7 disciplines). An example failure case is shown on the right.}
    \label{fig:paper_ask_component}
    \vspace{-1em}
\end{figure}

Existing evaluations fail to capture the full scholarly workflow. 
Most hallucination studies test LLMs without search augmentation~\cite{spylab2025hallucinations,buchanan2025hallucination}, showing that fabricated citations often originate from outdated internal knowledge. 
Document understanding benchmarks~\cite{dasigi2021qasper} evaluate whether models can answer questions about pre-collected papers rather than discovering them. 
Agentic systems are typically assessed using custom retrieval tools~\cite{openai2025whyhallucinate,wu2025agentic}, not through the standard web interfaces (e.g., ChatGPT) that researchers use in practice.

In this work, we introduce \textbf{PaperAsk}, a benchmark that evaluates the reliability of LLMs on basic paper search and reading tasks through systematic human assessment. Our design philosophy prioritizes realism over scale: we evaluate commercial LLMs (GPT-4o, GPT-5, and Gemini 2.5 Flash) via \textit{web-based interfaces} where LLMs perform web search autonomously, without assumptions about paper availability or pre-processed inputs. PaperAsk covers four core operations—\textit{citation retrieval}, \textit{content extraction}, \textit{paper discovery}, and \textit{claim verification}—that any dependable research assistant should perform reliably. Tasks range from retrieving BibTeX entries for a paper to counting figures in specified documents and assessing whether papers address stated research topics and methods.

Our experiments reveal systematic failures across both LLMs and tasks. Citation retrieval performance degraded sharply—from 2–9\% failure when retrieving \textit{three papers per query} to 48–98\% at \textit{ten papers per query}. Content extraction failed in 72–91\% of cases, with models frequently returning abstract text when prompted for introductions. Paper discovery achieved F1 scores below 0.32, missing over 60\% of relevant literature.
The most revealing task was claim verification, which exposed how \emph{deployment architecture directly influences reliability}. When provided with paper URLs through web interfaces, models failed 12–18\% of the time, whereas the same models accessed via API calls with full-text availability failed only 1–3\%—a gap of 9–15 percentage points. Our manual inspection uncovers the root cause: when given a few specific paper URLs, the LLM-integrated search functionality tends to retrieve far more sources—often several times the requested number—polluting the context with conflicting information. GPT-5 typically refused to answer under such overload, whereas Gemini produced fabricated completions. Both behaviors arose from the same architectural trade-off: \textbf{the web services of these LLMs prioritize responsiveness through shallow, broad retrieval at the cost of accurate and focused content access.}

In summary, our main contributions are:
\begin{itemize}[topsep=2pt,itemsep=12pt,parsep=0pt,leftmargin=*]

\item We present \textbf{PaperAsk}, a benchmark that evaluates the reliability of LLMs in scholarly tasks using domain-agnostic queries executed through deployed web interfaces.

\item We conduct systematic human evaluations of frontier LLMs (GPT-4o, GPT-5, and Gemini 2.5 Flash), revealing distinct failure modes across tasks—including over-conservative refusals in multi-paper queries, fabricated metadata to complete responses, and extraction from incorrect document sections.

\item We characterize these failure patterns and develop a lightweight classifier that detects unreliable model outputs with 96\% accuracy, providing a practical filtering mechanism for real-world deployment.
\end{itemize}

\section{Related Work}

\subsection{LLMs as Research Assistants}

Academic search remains a foundational problem in information retrieval. Traditional approaches leverage citation networks and keyword matching~\cite{giles1998citeseer,semanticscholar2025}, while neural citation recommendation applies collaborative filtering and representation learning~\cite{ebesu2017neural,mcnee2006citesight}. 
Recent systems unite large language models with search, marking a clear architectural shift. 
Platforms such as ChatGPT with Deep Research~\cite{openai2025deepresearch}, Google’s Gemini~\cite{google2025gemini}, and Consensus~\cite{consensus2025} pair retrieval with generation, which in principle should curb hallucinations seen in parametric-only models: when sources are cited, users perceive answers as grounded. 
Yet it remains uncertain whether these search-augmented LLMs reliably perform fundamental scholarly operations.

\subsection{Evaluation of LLM Reliability in Scholarly Tasks}

Existing evaluation approaches have advanced our understanding of specific failure modes and capabilities, yet a gap remains in assessing the complete workflow as deployed.

\textbf{Citation hallucination and accuracy.} Early work documented fabricated citations in parametric models~\cite{spylab2025hallucinations}, establishing that hallucination occurs at measurable rates. SourceCheckup~\cite{deng2025sourcecheckup} introduced automated verification for medical LLM responses, finding 50-90\% lack complete source support even in GPT-4o with web search, demonstrating that citation problems persist despite search integration. However, the evaluation focuses on claim-source alignment within medical domains.

\textbf{Reading comprehension benchmarks.} Question answering over scientific documents has been extensively studied. QASPER~\cite{dasigi2021qasper} provides 5,049 questions over 1,585 papers, while LitSearch~\cite{litsearch2024} offers 597 literature search queries. Comparative studies~\cite{hkust2024trust} examine tools like Scite, Elicit, and Consensus, revealing substantial performance variation. These benchmarks provide valuable insights into model comprehension capabilities but either assume papers are pre-identified or evaluate retrieval separately from understanding.

\textbf{Retrieval-augmented generation.} RAG evaluations have examined system performance on complex reasoning tasks~\cite{openai2025whyhallucinate,xu2024hallucination}, typically using offline corpora (e.g., Wikipedia snapshots) with custom retrieval tools rather than live commercial search integration. While informative for understanding model capabilities under controlled conditions, this approach does not reveal how proprietary deployment mechanisms affect reliability. Furthermore, complex tasks naturally tolerate some error, making it difficult to establish baseline expectations for fundamental operations.

\textbf{Evaluation methodology.} Recent work increasingly applies LLMs to evaluate other LLMs' outputs~\cite{li2024llms,szymanski2025limitations}, offering scalable alternatives to human annotation. However, LLM-as-judge approaches provide limited insight into failure mechanisms: they assess output quality but cannot determine whether models accessed appropriate sources, extracted from correct document sections, or followed structural constraints. Understanding these failure modes requires human inspection of the complete retrieval-to-generation pipeline.

\textbf{Our focus.} Prior evaluations use APIs with controlled retrieval or assume papers are pre-identified. We evaluate deployed LLMs as users encounter them: commercial search-augmented LLMs accessed through web interfaces where retrieval, content access, and generation operate as integrated black boxes. Through systematic human evaluation, we assess whether these models reliably perform fundamental scholarly operations—locating papers, extracting content, discovering relevant work, and verifying claims. This approach reveals how deployment mechanisms affect reliability in ways invisible to API-based or isolated capability testing.

\section{Reliability Evaluation Tasks}~\label{section_3_evaluation_task}

We evaluate the reliability of LLMs across four fundamental scholarly operations, grouped into two categories: \textbf{paper search} and \textbf{paper reading}. 
The former assesses whether LLMs can locate papers from the web source. 
In contrast, the latter examines whether LLMs can reason over a web collection of papers to identify relevant works and evaluate whether the retrieved papers align with the given topical descriptions. Details of each task are described in the following subsections.

\subsection{Paper Search Tasks}

Paper search tasks require LLMs to demonstrate basic retrieval ability during web browsing. 
To align with our objective of evaluating realism rather than sophistication or scale, we avoid ambiguous search queries and tasks that require deep understanding or analysis of paper content. 
Instead, we focus purely on locating and extracting information from academic papers. 
Accordingly, we design two subtasks: \textbf{citation retrieval} and \textbf{content extraction}.

Citation retrieval evaluates whether models can locate papers by their exact titles and generate accurate bibliographic information—the most fundamental operation a research assistant performs regularly. 
Content extraction assesses whether models can access and extract specific elements from identified papers, such as figure counts or designated sentences. 
Together, these operations establish the baseline for reliability: if models cannot consistently locate papers or extract visible content, the feasibility of more advanced scholarly assistance becomes doubtful.

\begin{figure}[!t]
    \centering
    \includegraphics[width=1\linewidth]{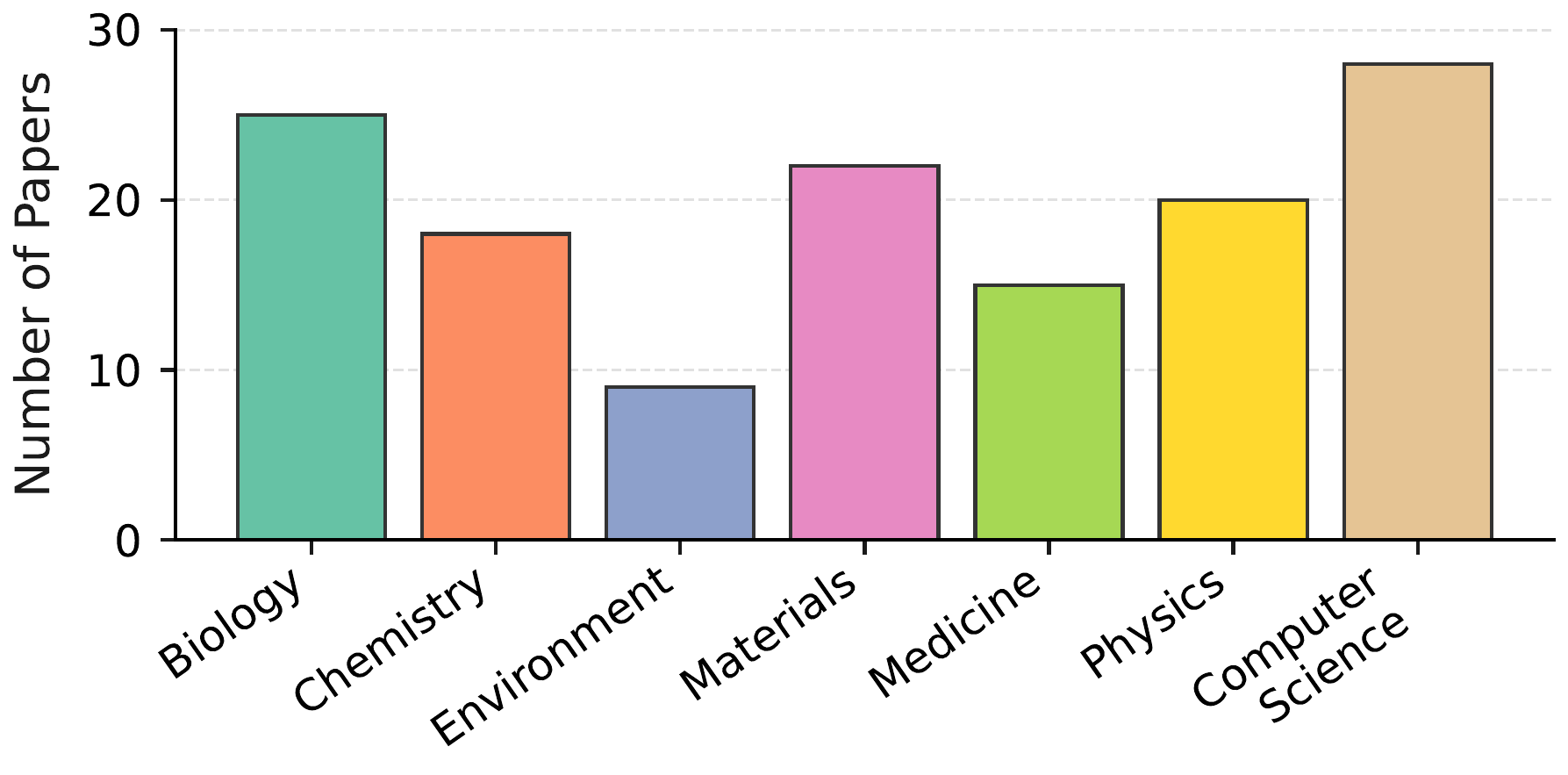}
    \caption{Distribution of papers across scientific fields.}
    \Description{}
    \label{fig:domain_fileds}
    \vspace{-0.3cm}
\end{figure}

\subsubsection{Citation Retrieval}
We sample 1000 papers published between 2022-2025 from arXiv's computer science domain, using their provided BibTeX as ground truth. From this pool, we construct 300 test queries by varying the number of papers requested per query, $n \in \{3,5,10\}$, where each query asks the LLM to return BibTeX entries for $n$ papers simultaneously. To ensure task solvability, we manually verify in common search engines (Google, Bing) that searching ``\texttt{\rgray{[{paper title}]} site:arxiv.org}'' returns the correct paper as the top result. We classify outputs as success (correct metadata with complete BibTeX fields) or failure otherwise. By selecting arXiv as our source, all papers are freely accessible, ensuring failures reflect system limitations rather than access restrictions.
\begin{claimbox}[Citation Retrieval]
\faRobot~You are a helpful assistant with arXiv search access.

\faSearch~\textbf{Query:} For the following paper titles: 
\begin{itemize}[leftmargin=*, noitemsep, topsep=2pt]
    \item Paper 1: \textit{``\rgray{Attention Is All You Need'}'}
    \item Paper 2: \textit{``\rgray{BERT: Pre-training of Deep Bidirectional Transformers}''}
    \item $\dots$
    \item Paper $n$: \textit{``\rgray{GPT-3: Language Models are Few-Shot Learners'}'}
\end{itemize}

\faTasks~\textbf{Task:} Return BibTeX entries for each paper.
\end{claimbox}

\subsubsection{Content Extraction}
Citation retrieval draws papers from computer science exclusively. To test whether extraction capabilities generalize across disciplines, we expand to diverse scientific domains. Specifically, we select 140 papers spanning physics, biology, mathematics, chemistry, medicine, economics, and computer science (Figure~\ref{fig:domain_fileds}), prioritizing published versions over preprints for consistency. Each query requests extraction from \textbf{a single paper}, unlike citation retrieval, where multiple papers are requested simultaneously. All papers have freely accessible PDFs online. Human annotators manually extract the five elements specified in the task—last sentence of the introduction, figure and table counts, and figure and table captions—establishing ground truth for evaluation.

\begin{claimbox}[Content Extraction Task]
\faRobot~You are a helpful assistant with search capabilities.

\faSearch~\textbf{Query:} Given a \rgray{[paper title \& URL]}, extract the following:
\begin{itemize}[leftmargin=*, noitemsep, topsep=2pt]
\item Last sentence of the introduction
\item Number of figures
\item Number of tables
\item Caption of the figure
\item Caption of the table
\end{itemize}

\faTasks~\textbf{Task:} Return all fields for the paper.
\end{claimbox}

\subsection{Paper Reading Tasks}
Paper reading tasks assess whether LLMs can comprehend research content deeply enough to judge relevance and verify claims.
They complement paper search tasks, which primarily test retrieval ability.
We propose two key sub-tasks: \textit{open-domain question answering} and \textit{claim verification}. The former evaluates whether models can identify papers that match given topic descriptions, while the latter tests whether models can accurately determine if provided papers address specific research claims.

However, evaluating paper reading capability presents a challenge in evaluation. Open-ended queries such as \textit{“find papers on neural scaling laws”} lack objective ground truth, as relevance judgments vary with researcher expertise, query interpretation, and task context. To mitigate this, we adopt queries from PaSa~\cite{he2025pasa}, a scholarly benchmark with expert-validated relevance labels. Moreover, we further constrain the search scope to the arXiv source within specified publication periods, enabling objective evaluation through exact matching on both topical and temporal criteria. This core idea trades generality for measurability—introducing controlled constraints to ensure reproducible and objective assessment while preserving the essence of the original task.

\subsubsection{Open-Domain Question Answering}
The query format follows: “Find papers on \rgray{[topic]}, published in \rgray{[month year]} on arXiv.”
Each topic is derived from the PaSa benchmark, which provides expert-validated ground-truth sets of relevant papers.
We additionally collect the publication date for each paper to ensure accurate temporal filtering during evaluation. A response succeeds only when it returns all ground truth papers without additions or omissions. We constructed 100 test queries, assessing exact match on arXiv metadata (e.g., authors, title, etc).

\begin{claimbox}[Open-Domain QA Task]
\faRobot~You are a helpful assistant with arXiv search access.

\faSearch~\textbf{Query:} Find papers on ``\rgray{Scaling Laws for Fine-Grained Mixture of Experts}'' published on \rgray{arXiv} in \rgray{February 2024}.

\faTasks~\textbf{Task:} Return BibTeX entries for all relevant papers.
\end{claimbox}

\subsubsection{Claim Verification}
Claim verification evaluates comprehension in an “open-book” setting.
We derive claims by reformulating open-domain QA queries into a binary verification format.
For example, the original query “\textit{What scaling laws govern mixture-of-experts models?}” requires both locating papers and synthesizing answers.
In the reformulated version, the LLMs is given a paper URL and a derived claim (e.g., “\textit{This \rgray{[paper with URL]} proposes scaling laws for mixture-of-experts}”) and determine whether the paper \texttt{supports/refutes} it.
Unlike open-domain QA, which is inherently ambiguous due to varying interpretations of topical relevance, claim verification provides a more constrained and objective setting.
It also preserves the need for in-depth content understanding, enabling us to assess the ability of LLMs to read research papers. Accordingly, we create each query that includes multiple claim–paper pairs ($n \in \{3, 5, 10 \}$), and success requires all pairs to be correctly verified, resulting in a total of 300 test cases.

\begin{claimbox}[Claim Verification Task ($n=2$)]
\faRobot~You are a claim verification assistant.

\textbf{Claim 1}: "This paper proposes methods based on LLMs and evaluates on HotpotQA."~\faLink~https://arxiv.org/abs/... \softgreen{(paper about LLMs and HotpotQA)}

\textbf{Claim 2}: "This paper presents research on robot decision making."~\faLink~https://arxiv.org/abs/... \softred{(paper about computer vision)}

\faTasks~\textbf{Task:} Return \texttt{SUPPORTED}/\texttt{REFUTED} with explanation.
\end{claimbox}

\subsection{Dataset Statistics}

Table~\ref{tab:task_overview} summarizes the evaluation scope. The benchmark comprises 840 test cases across four tasks, with sizes determined by task complexity and the need for reliable statistical assessment.

\begin{table}[h]
\centering
\caption{Overview of evaluation tasks and test sizes.}
\label{tab:task_overview}
\begin{tabular}{lc}
\toprule
\textbf{Task} & \textbf{Test Cases} \\
\midrule
Citation Retrieval & 300 \\
Content Extraction & 140 \\
Open-Domain QA & 100 \\
Claim Verification & 300 \\
\midrule
\textbf{Total} & \textbf{840} \\
\bottomrule
\end{tabular}
\end{table}

\begin{table*}[!ht]
\centering
\caption{Performance comparison of LLMs on PaperAsk.
Failure rate is reported for each task (lower is better). \textbf{$n$} indicates the number of papers, claims or questions per query.
\faDesktop~indicates an LLMs-based application with a built-in search function.}
\label{tab:main_results}
\renewcommand\arraystretch{1.32}
\begin{adjustbox}{max width=\textwidth}
\begin{tabular}{lcccc}
\Xhline{1.2pt}
\rowcolor{CadetBlue!23} 
\textbf{Category} & \textbf{Task}
& \faDesktop~\textbf{GPT-4o}
& \faDesktop~\textbf{GPT-5}
& \faDesktop~\textbf{Gemini 2.5 Flash} \\
\Xhline{1pt}

% ----- Paper Searching block -----
\rowcolor{CadetBlue!12}\multicolumn{5}{l}{\textbf{\textsc{Paper Searching}}}\\[-0.5ex]
\multirow{4}{*}{}
& Citation Retrieval ($n{=}3$)   & 2\%\redone{0\%}   & 9\%\red{7}  & 3\%\red{1}  \\
\rowcolor{gray!10} 
& Citation Retrieval ($n{=}5$)   & 18\%\red{5} & 26\%\red{13} & 13\%\redone{0\%} \\
& Citation Retrieval ($n{=}10$)  & 98\%\red{50} & 78\%\red{30} & 48\%\redone{0\%} \\
\rowcolor{gray!10} 
& Content Extraction             & 84.28\%\red{11.42} & 72.86\%\redone{0\%} & 91.42\%\red{18.56} \\[0.3ex]

% ----- Paper Understanding block -----
\rowcolor{CadetBlue!12}\multicolumn{5}{l}{\textbf{\textsc{Paper Reading}}}\\[-0.5ex]
\multirow{4}{*}{}
& Open-Domain QA                      & 73\%\red{5} & 80\%\red{12} & 68\%\redone{0\%} \\
\rowcolor{gray!10}
& Claim Verification ($n{=}3$)   & 18\%\red{6} & 12\%\redone{0\%} & 14\%\red{2} \\
& Claim Verification ($n{=}5$)   & 23\%\red{2} & 45\%\red{24} & 21\%\redone{0\%} \\
\rowcolor{gray!10}
& Claim Verification ($n{=}10$)  & 63\%\red{9} & 72\%\red{18} & 54\%\redone{0\%} \\

\Xhline{1.2pt}
\end{tabular}
\end{adjustbox}
\vspace{-0.6em}
\end{table*}

\section{Results and Analyses}

Having established the evaluation tasks, we now examine how current search-augmented LLMs perform on these fundamental scholarly operations. This section describes the evaluated systems, evaluation methodology, and results with analysis.

\subsection{Evaluated Search-Augmented LLMs}

We evaluate three widely adopted search-augmented LLMs: ChatGPT (GPT-4o and GPT-5) and Gemini 2.5 Flash. These systems represent the current state of commercially deployed research assistants. ChatGPT provides explicit search controls, which we enable to represent typical researcher usage. Gemini integrates search by default through Google Search grounding.

We manually input queries and collect complete responses. This preserves the full deployment architecture: proprietary retrieval mechanisms, ranking algorithms, and content access policies that differ from API behavior. Additionally, manual evaluation distinguishes our work from LLM-as-judge studies~\cite{szymanski2025limitations}, allowing for a detailed diagnosis of failure mechanisms. By directly examining the generated response, we identify failure patterns that automated metrics cannot detect.

\subsection{Evaluation Metrics}

Human annotators verify all responses against the ground truth established in Section~\ref{section_3_evaluation_task}. The primary metric is failure rate:
\[
\text{Failure Rate} = \frac{\text{Number of Failed Responses}}{\text{Total Number of Test Cases}} \times 100\%
\]

A response fails when it produces incorrect outputs or omits required information. Task-specific criteria are as follows:

\textbf{Citation Retrieval.} Failure occurs if any BibTeX entry contains incorrect metadata, missing required fields, or includes papers not in the query. For multi-paper queries ($n > 1$), success requires all $n$ citations to be correct.

\textbf{Content Extraction.} Failure occurs if any of the extraction items are incorrect or missing: last sentence of introduction, figure count, table caption, etc.

\textbf{Open-Domain QA.} Exact match on arXiv IDs and titles determines success. Failure occurs if ground truth papers are missing, timeframe constraints are violated, or incorrect papers appear. 

\textbf{Claim Verification.} Failure occurs if any verdict is incorrect. For a multi-claim query, success requires $n$ correct verdicts.

\begin{figure}[!ht]
    \centering
    \includegraphics[width=1\linewidth]{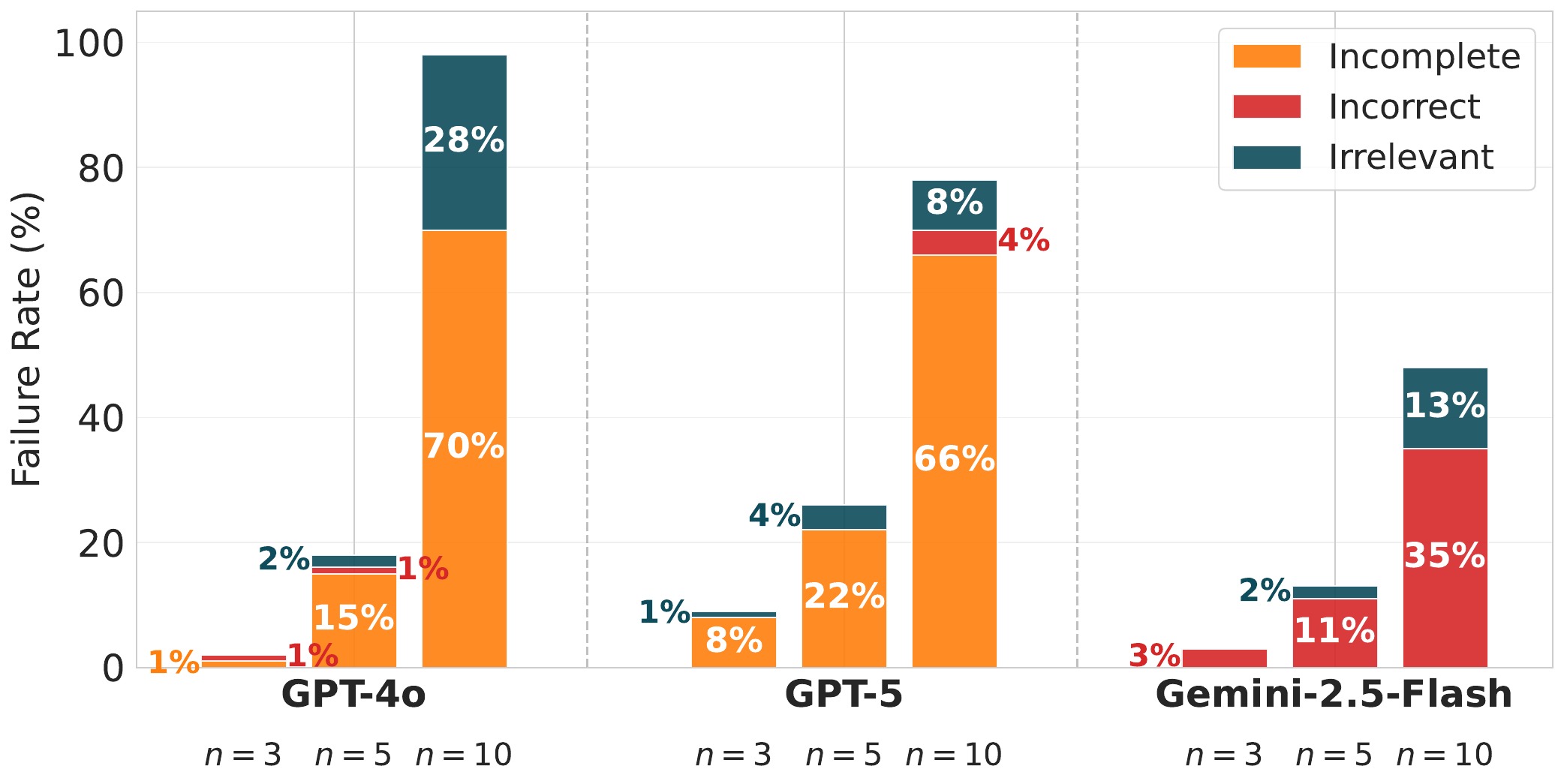}
\caption{Citation retrieval failure breakdown. Stacked bars show three failure types: incomplete responses (returning fewer citations than requested), incorrect metadata (mismatched titles/authors/arXiv IDs), and irrelevant citations (returning entries not from the arXiv source).}
    \label{fig:citation_failure}
    \vspace{-1em}
\end{figure}

\subsection{Result Summary}

Table~\ref{tab:main_results} shows pervasive failures across all fundamental scholarly tasks.\footnote{Evaluation results and test cases are available anonymously at \url{https://anonymous.4open.science/r/PaperAsk-26EE/} to support review.} Citation retrieval fails in 48-98\% of cases when requesting 10 papers simultaneously. Content extraction fails in 72-91\% of cases, with models consistently returning abstract text when asked for introduction sections. Open-domain QA fails in 68-80\% of cases, achieving F1 scores below 0.32 and missing over 60\% of relevant papers. Claim verification shows failure rates of 12-18\% when paper URLs are provided directly, rising to 54-72\% for multi-paper queries.

To better understand these failures, the following subsection discusses identified failure patterns for each task and highlights key observations \textbf{(Obs.)} supported by empirical evidence.

\subsection{Analyzing Citation Retrieval Failures}
\textbf{Obs.\ding{182} Requesting multiple papers simultaneously produces distinct failure modes across LLMs: ChatGPT refuses, while Gemini fabricates.} 
Although the requested papers can be easily located by simply searching their titles, which usually appear at the top of common search engines such as Google or Bing, LLMs still fail to retrieve the correct sources or extract their BibTeX entries. As shown in Figure~\ref{fig:citation_failure}, multi-paper queries exhibit a high failure rate. GPT-5 refuses 66\% of test cases and returns incomplete results, while the same papers succeed when queried individually ($n=1$).

We manually inspected the generated responses and cited sources returned by the LLMs and found that they intentionally expand their search scope beyond the specified constraints. 
Instead of searching for exact titles on the arXiv, LLMs show a strong preference for retrieving bibliographic information from multiple sources, including citations in GitHub repositories or Hugging Face pages.
This behavior inevitably introduces conflicting information from sources outside the intended domain.

Interestingly, LLMs respond to this noise through opposite strategies. ChatGPT responds conservatively: when retrieved content conflicts, it refuses to answer rather than risk errors, typically returning fewer than $n$ requested BibTeX entries. Gemini responds by fabricating: it always returns exactly $n$ BibTeX entries as requested with fabricated data (Table~\ref{tab:citation_examples}). At $n=10$, 35\% of responses contain fabricated arXiv IDs pointing to different papers. Gemini prioritizes appearing complete over being accurate.

Given the above observations, this pattern likely reflects reinforcement learning optimization, where language models learn to favor task completion over strict adherence to constraints~\cite{schulman2017ppo,christiano2017rlhf}. The LLMs tend to act helpfully, prioritizing responses that appear complete to maximize their perceived reward. However, it remains uncertain whether these failures stem from the concurrent processing of multiple papers or from deeper limitations in the underlying search mechanism. To investigate this, we next examine single-paper tasks to isolate the contributing factors.

% We verified: every requested paper ranks first in Google search. If basic search works, why do failure rates reach 98\% for ChatGPT at $n=10$?

\begin{table}[!t]
\centering
\caption{Citation retrieval failure example.}
\label{tab:citation_examples}
\begin{adjustbox}{max width=\columnwidth}
\small
\begin{tabular}{@{}p{1.8cm}p{5.2cm}@{}}
\toprule
\multicolumn{2}{c}{\textbf{\faDesktop~Gemini 2.5 Flash}} \\
\midrule
\textbf{Query} & ... For the following papers, use arXiv search and return BibTeX entries ...\\
& 1. In-context Pretraining... \\
& 2. Why Can GPT Learn In-Context?... \\
& 3. Transformers Learn Higher-Order... \\
\addlinespace[0.4em]

\textbf{Output} & \texttt{1. @article\{sun2024,...\}} \softgreen{\faCheckCircle~Correct}  \\
& \texttt{2. @article\{dai2022,...\}} \softgreen{\faCheckCircle~Correct}  \\
& \texttt{\softred{3. @article\{shi2024,...}}  \\
& \texttt{~~~~\softred{journal=\{arXiv:2407.06948\}\}}} \\
& \softred{\faTimesCircle~Wrong arXiv ID} \\
\addlinespace[0.4em]

\textbf{Error} & arXiv:2407.06948 points to a different paper.\\
\bottomrule
\end{tabular}
\end{adjustbox}
\vspace{-2em}
\end{table}

\subsection{Analyzing Content Extraction}
\textbf{Obs.\ding{183} Failures persist in single-paper tasks: models prioritize text similarity over task instruction, assembling responses from multiple sources based solely on semantic relevance.}

As shown in Table~\ref{tab:main_results}, failure rates reach 72-91\% across LLMs, with no observable correlation to scientific fields (Figure~\ref{fig:domain_fileds}). The primary error pattern involves locating the last sentence of the introduction section. LLMs consistently return abstract content instead, as shown in Figure~\ref{fig:paper_ask_component}. This aligns with our earlier observation: abstracts frequently appear in shallow search result snippets, making them more accessible than full paper content. In contrast, counting figures and tables produces relatively lower error rates, initially suggesting LLMs are still capable of accessing complete paper content when performing the web search. 

However, manual inspection of caption extraction reveals otherwise. Erroneous captions do not appear in the target paper's PDF. We collected generated responses and examined cited sources, finding that incorrect text originates from related but non-target webpages. For example, a blog post mentioning the requested paper might paraphrase a table caption. LLMs treat this paraphrased text as the actual caption, \textbf{assembling answers by cross-referencing multiple sources for each extraction question}.

Additionally, GPT-5 exhibits high confidence in this task despite its frequent refusals in citation retrieval (i.e., incomplete failures). This behavior exposes a fundamental limitation: LLMs cannot reliably distinguish task instructions from retrieved content, treating any semantically relevant text as valid information regardless of source. Consequently, this makes indirect prompt injection a critical attack surface~\cite{yi2025benchmarking, greshake2023not}. Attackers exploit this by injecting poisoned text into publicly editable sources~\cite{carlini2024poisoning}. During web search, LLMs incorporate this malicious content, generating attacker-predefined responses.

Moreover, we observe an \textbf{intriguing pattern for well-known papers}: LLMs persist with fixed responses rather than adapting to retrieved content. As shown in Table~\ref{tab:content_extraction_example}, LLMs provide incorrect answers regardless of what appears in target documents, even when the retrieved information contradicts their responses.

\begin{figure}[!t]
    \centering
    \includegraphics[width=1\linewidth]{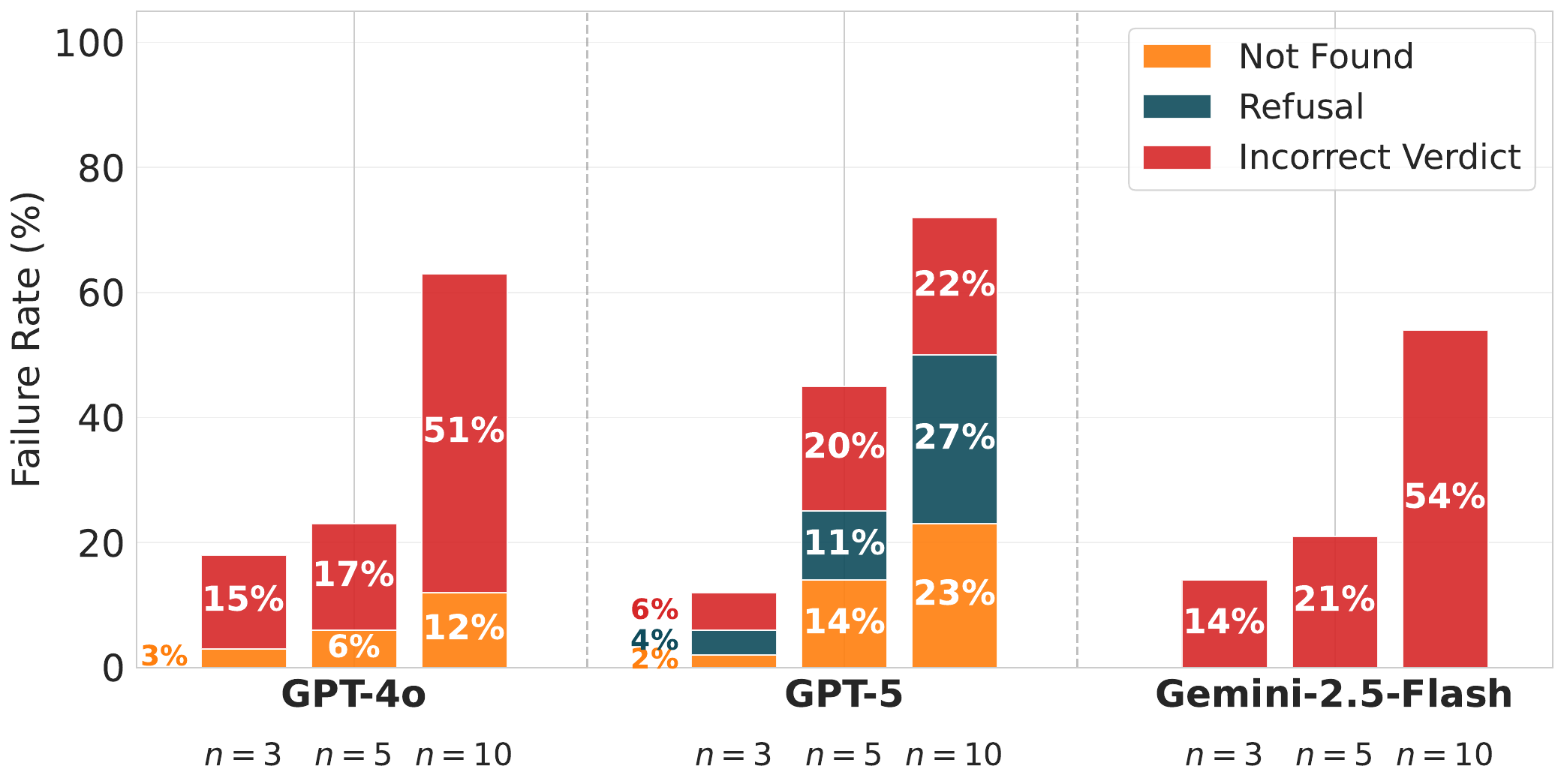}
    \caption{Claim verification failure type distribution across query sizes. Three failure types: not found (LLMs claims inability to locate provided paper), refusal (LLMs refuses to verify due to uncertainty), and incorrect verdict (wrong classification of claim).}
    \Description{Bar chart showing failure type distribution across query sizes: not found, refusal, and incorrect verdict.}
    \label{fig:factchecking_failure}
    \vspace{-1em}
\end{figure}
\begin{table}[!t]
\centering
\caption{Content extraction example.}
\label{tab:content_extraction_example}
\begin{adjustbox}{max width=\columnwidth}
\small
\begin{tabular}{@{}p{1.8cm}p{5.2cm}@{}}
\toprule
\multicolumn{2}{c}{\textbf{\faDesktop~GPT-5}} \\
\midrule
\textbf{Query} & \faSearch~For ``Attention Is All You Need'' (\textbf{ACM NIPS version}), count number of tables and provide full caption for last table. \\
\addlinespace[0.4em]
\textbf{Response} & \rorange{\faQuestionCircle~I checked the NIPS version.} There are 4 tables in total. Table 4 caption: \rorange{``The Transformer generalizes well to English constituency parsing (Results are on Section 23 of WSJ)''} \\
\addlinespace[0.4em]
\textbf{Ground Truth} & \faFilePdf~\textbf{NIPS:}$^1$ \softred{Only 3 tables (no Table 4).} \\
\bottomrule
\end{tabular}
\end{adjustbox}
\noindent\small{$^1$\url{https://dl.acm.org/doi/pdf/10.5555/3295222.3295349}}
\vspace{-1em}
\end{table}

\subsection{Analyzing Open-Domain QA Failures}
Open-domain QA tests discovery capability: identifying papers from topical descriptions without knowledge of the existing papers.

\textbf{Obs.\ding{184} Search relies on shallow keyword and fuzzy matching with limited coverage.} Results are shown in Tables~\ref{tab:main_results}, \ref{tab:retrieval_performance}, and \ref{tab:error_breakdown}. All LLMs achieve F1 scores below 0.32. Recall stays under 40\%, meaning LLMs miss over 60\% of relevant papers. Precision remains under 30\%, meaning over 70\% of returned papers are incorrect. GPT-5 achieves 33\% recall and 14\% precision, missing two-thirds of relevant papers while returning 86\% irrelevant content.

Manual inspection reveals distinct search behaviors. ChatGPT exhibits rigid keyword-based matching. Responses frequently state "I could not find any results matching keywords," indicating inflexible query processing. GPT-5 attempts to compensate by relaxing temporal constraints to find topic-matching papers, achieving lower topic mismatch (38\% vs GPT-4o's 43\%) but similar date violations (44\%). GPT-4o's returned papers are often topically irrelevant, suggesting keyword matching fails to capture semantic meaning.

Gemini demonstrates broader but noisier search. It achieves the lowest date mismatch (30\%) and competitive topic matching (42\%), indicating stronger search capabilities. However, source mismatch reaches 28\%. Gemini's multi-source retrieval extends beyond arXiv to conferences, workshops, and preprint servers. This explains the high incorrect metadata rates observed in Figure~\ref{fig:citation_failure}: noise stems from searching broadly across diverse sources.

\textbf{Obs.\ding{185} LLMs judge relevance from search snippets rather than complete documents.} Table~\ref{tab:openqa_example} illustrates this mechanism. GPT-4o initially returns an incorrect paper by matching only title keywords such as “Diffusion”. When explicitly instructed to “search content and verify,” the LLMs recognize the mismatch and retrieve the correct paper. Our analysis of successful responses suggests that papers are retrieved primarily when their titles or abstracts contain phrases that closely match the topical description.

\begin{table}[!t]
\centering
\caption{Retrieval performance in open-domain QA.}
\label{tab:retrieval_performance}
\begin{adjustbox}{max width=\columnwidth}
\begin{tabular}{l|ccc}
\Xhline{1.2pt}
\textbf{Model} & \textbf{Recall (\%)} & \textbf{Precision (\%)} & \textbf{F1} \\
\Xhline{1pt}
GPT-4o & 32.5 & 22.0 & 0.26 \\
GPT-5 & 33.33 & 14.29 & 0.20 \\
Gemini 2.5 Flash & 38.0 & 28.0 & 0.32 \\
\Xhline{1.2pt}
\end{tabular}
\end{adjustbox}
\vspace{-0.14cm}
\end{table}

\begin{table}[!t]
\centering
\caption{Error type distribution among incorrect responses.}
\label{tab:error_breakdown}
\begin{adjustbox}{max width=0.95\columnwidth}
\begin{tabular}{l|ccc}
\Xhline{1.2pt}
\textbf{Model} & \textbf{Date} & \textbf{Topic} & \textbf{Source} \\
& \textbf{Mismatch (\%)} & \textbf{Mismatch (\%)} & \textbf{Mismatch (\%)} \\
\Xhline{1pt}
GPT-4o & 45 & 43 & 12 \\
GPT-5 & 44 & 38 & 18 \\
Gemini 2.5 Flash & 30 & 42 & 28 \\
\Xhline{1.2pt}
\end{tabular}
\end{adjustbox}
\vspace{-0.2cm}
\end{table}

\begin{table}[!t]
\centering
\caption{Open-domain QA example showing search failure with intact judgment.}
\label{tab:openqa_example}
\begin{adjustbox}{max width=\columnwidth}
\small
\begin{tabular}{@{}p{1.8cm}p{5.2cm}@{}}
\toprule
\multicolumn{2}{c}{\textbf{\faDesktop~GPT-5}} \\
\midrule
\textbf{Query} & You are a research assistant with arXiv search... User asks: ``protect generation quality of LLMs under vocabulary watermarking'', list all papers published in August 2023. \\
\addlinespace[0.4em]
\textbf{Response} & \softred{\faTimesCircle~I did not find any papers published in August 2023 that matched the \textbf{keywords}.} \\
\addlinespace[0.4em]
\textbf{Follow-up} & \faQuestionCircle~How about ``Advancing Beyond Identification: Multi-bit Watermark for LLMs''? Does it fit? \\
\addlinespace[0.4em]
\textbf{Response} & \softgreen{\faCheckCircle~Yes, this paper fits well.} Focuses on embedding multi-bit information while maintaining text quality. \\
\addlinespace[0.4em]
\toprule
\multicolumn{2}{c}{\textbf{\faDesktop~GPT-4o}} \\
\toprule
\textbf{Query} & You are a research assistant with arXiv search... User asks: study about utilizing reinforcement learning to optimize diffusion models for video generation, list all papers published in December 2023. \\
\addlinespace[0.4em]
\textbf{Response} &  \texttt{\softred{3. @article\{huang2023,...}}  \\
& \texttt{\softred{title=\{Diffusion Reward: Learning Rewards via Conditional Video Diffusion\}\}}} \\ \\
\addlinespace[0.4em]
\textbf{Follow-up} & \faQuestionCircle~Please search content and verify. \\
\addlinespace[0.4em]
\textbf{Response} & \softred{\faTimesCircle~No, it does not primarily use RL to optimize video diffusion models...}. However, this is a more relevant paper. \softgreen{\faCheckCircle~Correct paper return...} \\

\bottomrule
\end{tabular}
\end{adjustbox}
\end{table}

\subsection{Analyzing Claim Verification}
Claim verification eliminates search complexity by providing paper URLs directly. This creates an “open-book” setting where LLMs have prior knowledge of the existing paper source.

\textbf{Obs.\ding{186} Retrieval depth limits performance.} As shown in Table~\ref{tab:main_results}, failure rates vary significantly across LLMs. At $n=3$ claim-paper pairs: GPT-5 fails at 12\%, GPT-4o at 18\%, Gemini at 14\%. At $n=10$ pairs: GPT-5 jumps to 72\%, Gemini reaches 54\%.

We observe LLMs follow a \textbf{sequential reading pattern}: examining abstracts first, then accessing full content only when initial information proves insufficient.
For example, when relevant details appear early in the abstract or introduction, the LLMs stop reading and generate an answer. In contrast, if the key idea is introduced later in the paper, the LLMs often fail to continue reading, leading to incomplete or incorrect responses. To validate this observation, we provide the complete paper content to LLMs through the API. Table~\ref{tab:api_factcheck} shows API access with full text fails at only 1-3\%, compared to web-interface LLM usage
 failing at 12-18\% for $n=3$—a 9-17 percentage point improvement. This suggests incompatibility between web search and LLM reasoning: constrained retrieval pressures LLMs toward premature conclusions using approximate answers to save computational overhead.

\textbf{Obs.\ding{187} Web search retrieves far more sources than specified.}
Manual investigation through ChatGPT's source references (\faGlobe~icon) reveals a second incompatibility: web search expands beyond provided URLs, often retrieving multiple times the requested number of sources. This excessive retrieval creates context pollution, overwhelming LLMs' ability to isolate requested papers. In practice, we observed that whenever a query triggers a search operation, LLMs consistently retrieve information from multiple sources rather than focusing on the specified one.

\begin{table}[!t]
\centering
\caption{Comparing claim verification performance: Web-based (black-box) vs API (full text provided) at $n=3$.}
\label{tab:api_factcheck}
\renewcommand\arraystretch{1.2}
\begin{adjustbox}{max width=\columnwidth}
\begin{tabular}{l|cc}
\Xhline{1.2pt}
\textbf{Model} & \faDesktop~Web-based (\%) & \faCode~API (\%) \\
\Xhline{1pt}
GPT-4o &  18 & 3 \\
GPT-5 & 12 & 1 \\
Gemini 2.5 Flash & 14 & 3 \\
\Xhline{1.2pt}
\end{tabular}
\end{adjustbox}
\vspace{-1.3em}
\end{table}

\section{Ablation Studies}

Our evaluation reveals systematic failures in how commercial LLMs interact with black-box web search. To isolate and assess LLMs’ advanced reasoning capabilities independently, we conduct controlled experiments using agent-based workflows with full observability (i.e., API-based testing).

\subsection{Experimental Setup}

We evaluate content extraction tasks using three model configurations: GPT-4o (non-reasoning baseline), GPT-5 with minimal reasoning (default), and GPT-5 with extended reasoning budgets. Each model operates under two tool environments:

\begin{itemize}[leftmargin=*, noitemsep, topsep=2pt]
    \item \textbf{Integrated Search:} OpenAI's native web search tool combines retrieval and scraping.
    \item \textbf{Customized Search:} We adopt Google Serper API with separate retrieval and fetch functions. Retrieval returns top-5 snippet results with URLs. Fetch retrieves full-page content given a URL.
\end{itemize}
We adopt the OpenAI Agent SDK~\cite{openai2024agents} to simulate the search-augmented LLMs. Agents follow observe-then-act workflows with a maximum of 10 tool interactions per query. We classify responses into four categories: correct (all extracted items match ground truth), error (incorrect extraction), reject (LLMs acknowledges failure), and timeout (exceeds interaction limit without producing output).

\subsection{Results}

\begin{figure}
    \centering
    \includegraphics[width=0.9\linewidth]{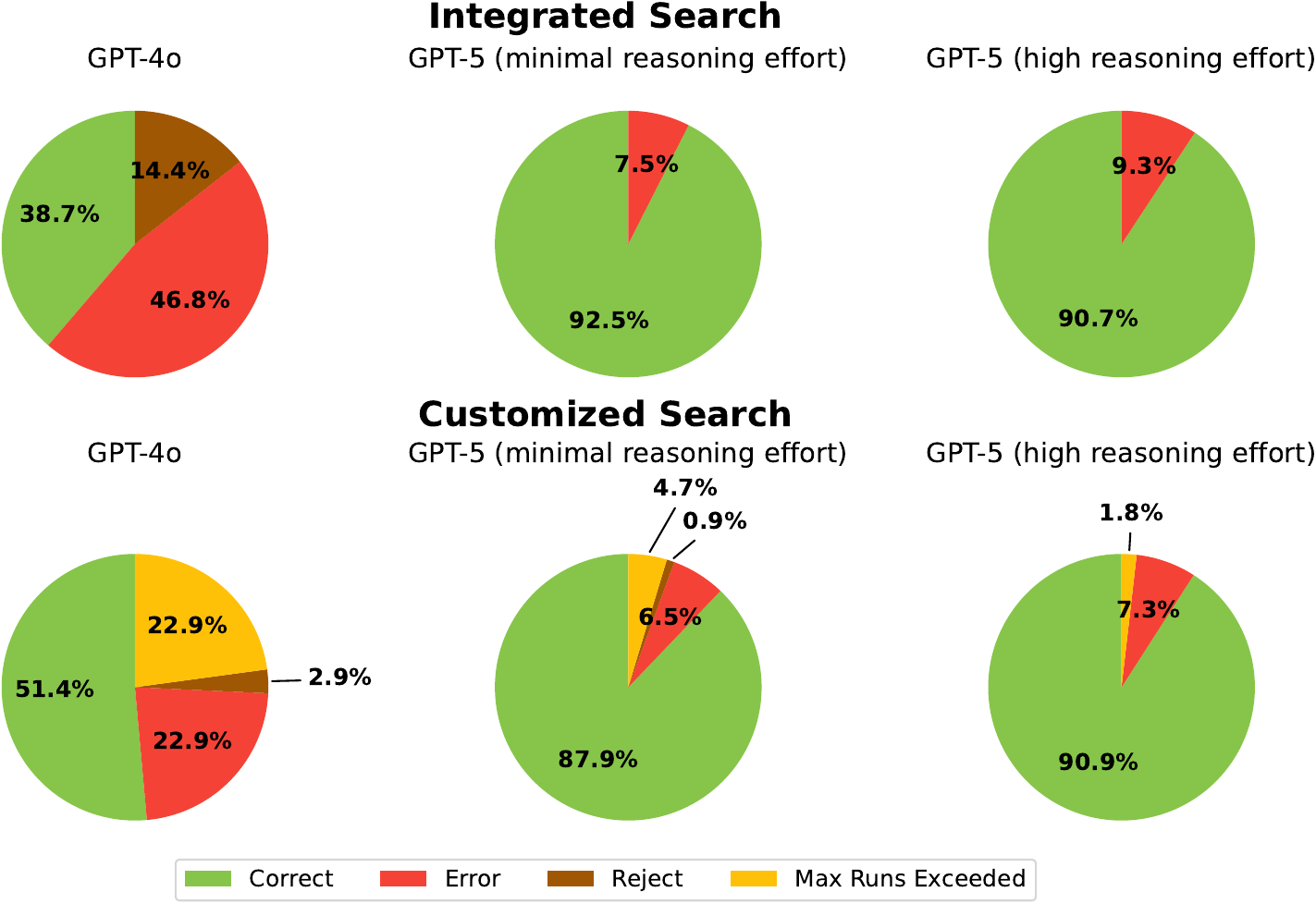}
    \caption{Agent performance on content extraction comparing GPT-5 (with high/minimal reasoning) and GPT-4o (non-reasoning) across two tool configurations: OpenAI's integrated search versus custom web search.}
    \label{fig:experiment-1-1}
    \vspace{-1em}
\end{figure}

Figure~\ref{fig:experiment-1-1} compares GPT-5 (reasoning) against GPT-4o (non-reasoning baseline). Reasoning capability creates larger gaps than tool configuration. Accuracy ranges 88-92\% for reasoning versus 39-51\% for baseline. Errors show the same pattern: 7-9\% versus 23-47\%, with baseline models refusing 14-23\% of queries.

Custom tools improve GPT-4o by 13 percentage points but GPT-5 by only 5, indicating that models with stronger internal reasoning rely less on external retrieval. When retrieval provides limited context, GPT-4o often fails, whereas GPT-5 compensates through iterative verification. The gap narrows with complete access but never fully closes, suggesting that architectural constraints affect models differently depending on their reasoning capacity.

Notably, controlled environments reduce performance gaps between LLMs compared to regular user-facing deployments. Such systems often trade accuracy for usability, using shallow retrieval and limited reasoning budgets. Simulated agent environments remove these constraints but incur response times that are 3–5 times longer and significantly higher computational costs.

\subsection{Analysis}

These results reveal fundamental challenges in combining search functions with modern LLMs. Even a simple query can trigger exponential growth in both reasoning and retrieval complexity. Current search mechanisms lack native compatibility with reasoning-driven workflows. As a result, LLMs inevitably explore multiple candidate sources before validating correctness, causing a combinatorial explosion in the retrieved context.

\section{How to Improve Reliability}
\begin{figure}[!t]
    \centering
    \includegraphics[width=0.9\linewidth]{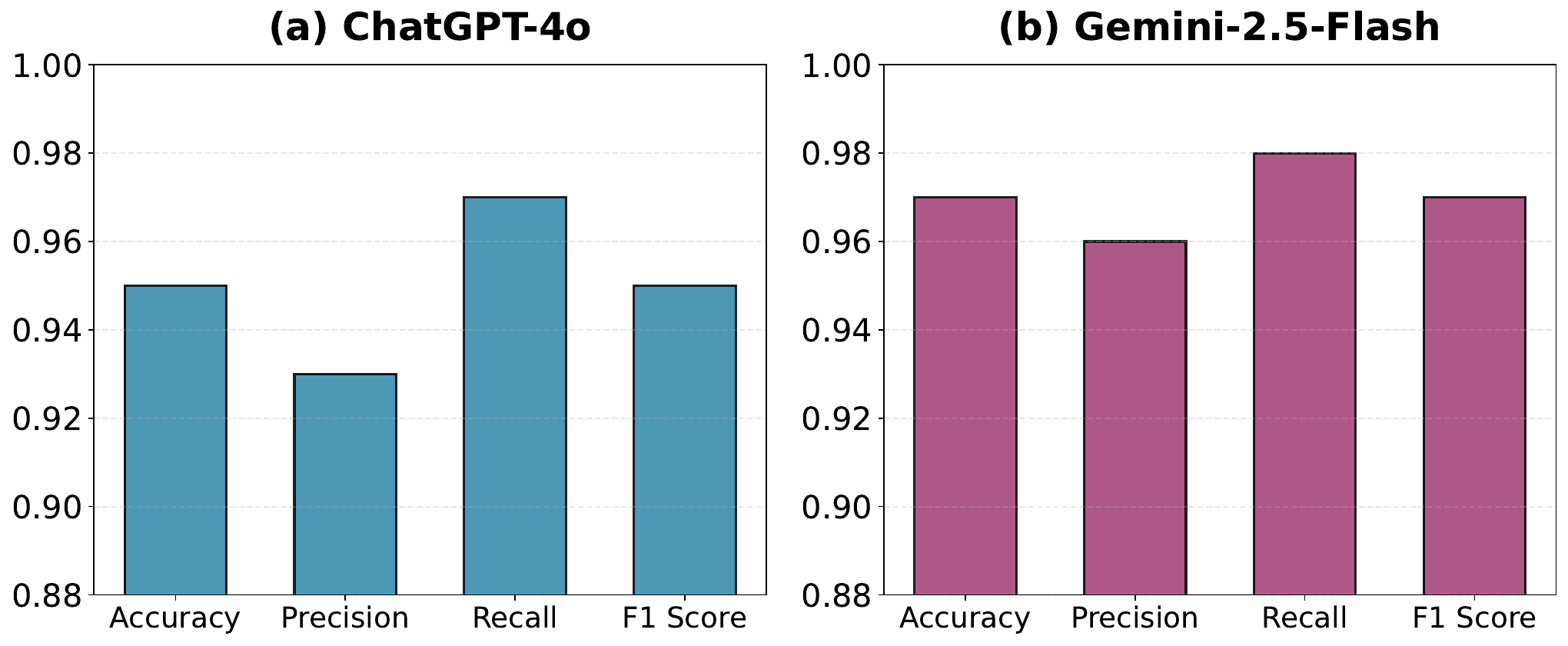}
    \caption{Performance of Llama-3.1-8B-Instruct classifier trained on annotated explanations from ChatGPT-4o and Gemini-2.5-Flash. The classifier demonstrates consistent high performance across accuracy, precision, recall, and F1 metrics when evaluated on held-out test data, validating its effectiveness as a lightweight reliability filter.}
    \label{fig:classifier}
    \vspace{-1.5em}
\end{figure}

Existing work addresses LLM retrieval limitations and hallucination through rerankers, RAG systems, and multi-agent architectures~\cite{tan2024small,zhang2025agent,gao2025llm4rerank}. However, these approaches introduce substantial computational overhead. Our work evaluates LLM capabilities on scholarly tasks from the perspective of public users. Therefore, we propose simple and effective methods to improve reliability without requiring complex infrastructure.

\textbf{Method \ding{182}: Simplify queries to single tasks.} Multi-question queries trigger exponential search space growth. Instead of requesting multiple questions simultaneously, decompose them into atomic requests: one paper per citation retrieval query and one question per paper for content extraction. This reduces context pollution and prevents models from confusing information across sources.

\textbf{Method \ding{183}: Disable search function for reasoning models.} For ChatGPT, turning off the built-in search function paradoxically improves accuracy. Models still exhibit search behavior through extended reasoning but execute more thoroughly. Response times increase 3-5 times, but correctness improves as models avoid shallow retrieval constraints imposed by the search function.

\textbf{Method \ding{184}: Deploy lightweight reliability classifiers.} For each task in PaperAsk, we prompt LLMs to provide explanations for their responses. We collect these explanations from both correct and failure cases, then annotate them with ground truth labels (correct/failure). We split the annotated data 80/20 for training and testing. Using the training set, we fine-tune Llama-3.1-8B-Instruct to classify response reliability. The classifier learns to detect hallucination signals from explanation patterns: fabricated metadata justifications, semantic similarity violations, and structural inconsistencies in reasoning. Figure~\ref{fig:classifier} shows evaluation on the held-out test set, where the classifier achieves 96\% overall accuracy in detecting response correctness, with F1 scores ranging from 0.95 to 0.97. This post-processing filter flags unreliable outputs without requiring architectural changes to deployment systems.

\section{Conclusion}
In this work, we introduced \textbf{PaperAsk}, a benchmark that evaluated search-augmented LLMs on scholarly tasks. Systematic human evaluations revealed pervasive reliability failures across citation retrieval, content extraction, paper discovery, and claim verification, with error rates approaching near-total failure on fundamental operations. These failures followed consistent patterns: LLMs tended to prioritize semantic similarity over structural fidelity, while integrated web search retrieved shallow snippets that polluted context with conflicting information. Controlled experiments showed that granting access to complete documents improved reliability by an order of magnitude, indicating that deployment architectures constrained otherwise capable models more severely than their reasoning limitations did. Future work could extend this evaluation to paywalled content and complex reasoning tasks, design adaptive retrieval mechanisms that adjust depth according to task requirements, and establish standardized protocols for assessing evolving deployed systems. PaperAsk offered a reproducible framework for measuring progress toward scholarly assistance systems that enhance—rather than hinder—reliable research workflows.

\bibliographystyle{ACM-Reference-Format}
\bibliography{custom}

\end{document}